# *Grain Selection Growth of Soft Metal in Electrochemical Processes*

Minghao Zhang[1,+], Karnpiwat Tantratian[2,+], So-Yeon Ham[3], Zhuo Wang[4], Mehdi Chouchane[1], Ryosuke Shimizu[5], Shuang Bai[1], Hedi Yang[1], Zhao Liu[6], Letian Li[6], Amir Avishai[7], Lei Chen[4, *], Ying Shirley Meng[1, 5, *]

1. Pritzker School of Molecular Engineering, The University of Chicago; Chicago, IL 60637, USA.
2. Department of Metallurgical Engineering, Chulalongkorn University; Bangkok, 10330, Thailand.
3. Materials Science and Engineering Program, University of California San Diego; La Jolla, CA 92093, USA.
4. Department of Mechanical Engineering, University of Michigan-Dearborn; Dearborn, MI 48128, USA.
5. Department of NanoEngineering, University of California San Diego; La Jolla, CA 92093, USA.
6. Materials and Structural Analysis, Thermo Fisher Scientific; Hillsboro, OR 97124, USA.
7. The Core Center of Excellence in Nano Imaging, University of Southern California; Los Angeles, CA 90089, USA.

[+]These authors contributed equally to the work.

*Corresponding author: leichn@umich.edu, shirleymeng@uchicago.edu

## Summary

Soft metals like lithium and sodium play a critical role in battery technology owing to their high energy density. Texture formation by grain selection growth of soft metals during electrochemical processes is a crucial factor affecting power and safety. Developing a framework to understand and control grain growth is a multifaceted challenge. Here, a general thermodynamic theory and phase-field model are formulated to study grain selection growth of soft metals. Our study focuses on the interplay between surface energy and atomic mobility-related intrinsic strain energy in grain selection growth. Differences in grain selection growth arise from the anisotropy in surface energy and diffusion barrier of soft metal atoms. Our findings highlight the kinetic limitations of solid-state Li metal batteries, which originate from load stress-induced surface energy anisotropy. These insights lead to the development of an amorphous $Li_xSi_{1-x}$ (0.50<x<0.79) seed layer, improving the critical current density at room temperature for anode-free Li solid-state batteries through the control of grain selection growth.

# Introduction

The transition from intercalation-type anodes to metallic anodes represents a significant paradigm shift in battery technology.[1] Li metal is considered an ultimate anode material for future high-energy rechargeable batteries with specific energy higher than 350 Wh/kg if paired with intercalation cathodes, and 500 Wh/kg if paired with conversion cathodes.[1] The energy density of Li metal batteries to withstand repeated charge and discharge cycles depends on the efficiency of lithium deposition and stripping. The morphology and microstructure of deposited lithium metal is a critical factor influencing the Coulombic efficiency (CE) and cycle life of Li metal batteries.[2,3] The ideal microstructure for Li deposits entails dense formations with minimal porosity (< 1%), a columnar structure featuring reduced surface area, and large grain sizes (>50 μm) exhibiting uniform defect distribution.[4] These favored attributes promote uniform Li stripping at the reaction front, thereby avoiding the formation of highly porous and whisker-like inactive Li structures.

In the Li metal battery with solid-state electrolytes (SSE) such as $Li_7La_3Zr_2O_{12}$ (LLZO) and $Li_6PS_5Cl$ (LPSCl), electrochemically deposited lithium metal typically exhibits a fully dense morphology with large grain size.[5,6] However, low critical current densities (<1.5 mA/cm$^2$) are reported over which a cell failure occurs (**Fig. S1** and **Table S1**). While elevated temperatures yield high current densities (~3 mA/cm$^2$), such values remain incomparable to those achieved by lithium metal batteries with liquid electrolytes at room temperature,[7] as demonstrated in **Fig. S1**. The kinetic limitations of Li metal are fundamentally influenced by crystallographic orientation, owing to the anisotropic nature of Li metal growth. Unlike the characteristic morphology, the orientation of Li metal growth with the SSE has not been previously explored.

Metals growth can exhibit a preferred crystallographic orientation, known as texture. The variations of total energy during deposition consist of changes in strain, thermal, and surface energy density, $\Delta U = \Delta \Gamma_{surface} + \Delta F_{strain} + \Delta F_{thermal}$.[8,9] Grain selection growth, as described by thermodynamic theory, is a fundamental process that minimizes the total energy of the system. Generally, grains characterized by low surface energy are thermodynamically favored.[10] Furthermore, grains with higher atomic mobility are preferable as rapid movement of atoms facilitate interface migration, inducing less strain energy.[11] Additionally, grains with high thermal conductivity grow preferentially as they enable rapid heat dissipation.[12] However, when determining texture, complexities arise because grains with the lowest surface energy may not invariably exhibit the lowest strain energy and thermal energy.

In the present study, we formulate a thermodynamically consistent theoretical framework for deposited soft metals demonstrating texture. Given the thin nature of deposited film (<50 μm), heat generated during grain growth efficiently dissipates to the surface, thus the thermal anisotropy can be disregarded. As a result, the competition is primarily between surface energy and strain energy. Considering two island grain nuclei, as they grow, they naturally close the gap between them to reduce total energy, transitioning from two free surface energies to one grain boundary energy (**Fig. 1**). The grain selection growth is thermodynamically controlled by the energy density difference between grain 1 and grain 2,

$$\Delta U_{12} = \Delta \Gamma_{surface} + \Delta F_{strain} \qquad \text{(Eq. 1)}$$

where $\Delta \Gamma_{surface} = (E_{S,1} - E_{S,2})/h$ is the surface energy density difference, $h$ is the film thickness (m), and $E_{S,i}$ is the surface energy of grain $i$ (J/m²). The difference in strain energy density, $\Delta F_{strain}$, arises from both intrinsic and extrinsic stresses.

The extrinsic strain depends on external stimuli such as applied stress ($\sigma$), rather the mechanical properties.[13,14] Extrinsic strain arises from external stimuli, such as stacking pressure in anode-free solid-state batteries. As typical ranges of applied stacking pressure usually exceed yield strength of soft metals, creep behavior is expected. However, the impact of creep anisotropy is expected to be small compared to other factors, such as surface energy and self-diffusion barriers, and can thus be considered negligible. Meanwhile, intrinsic stress is closely related to atomic mobility. As two grains close the gap to reduce surface energies, they generate tensile stress within. However, a higher atomic mobility facilitates such interface movement through atomic migration, rather than mechanical deformation alone, resulting in less tensile, or even compressive strain (**Fig. 1**). The relationship between intrinsic strain and atomic mobility is

$$\varepsilon_{intrinsic} = \varepsilon_c + (\varepsilon_T - \varepsilon_c) exp\left(\frac{-\beta D_i}{LR}\right) \qquad \text{(Eq. 2)}$$

where $D_i$ is the self-diffusion coefficient of each grain.[15] $\varepsilon_T$ is the tensile strain induced when grains mechanically deforming to close the gap without atomic diffusion assistance. $\varepsilon_c$ is the compressive strain from atomic additions to the grain boundary. $\beta$ is a fitting constant. $L$ and $R$ is grain size and the deposition rate, respectively.

Recent first-principles calculation results suggest that self-diffusion barriers for Li, Na, and K are strongly anisotropic. In addition, surface energy also exhibits anisotropy, and can significantly increase when the lattice parameter shifts due to load stress (~ 3% strain).[16] To capture the grain selection, we implement the thermodynamic theory into grain growth phase-field model for electrochemically deposited soft metals within the context of SSE.

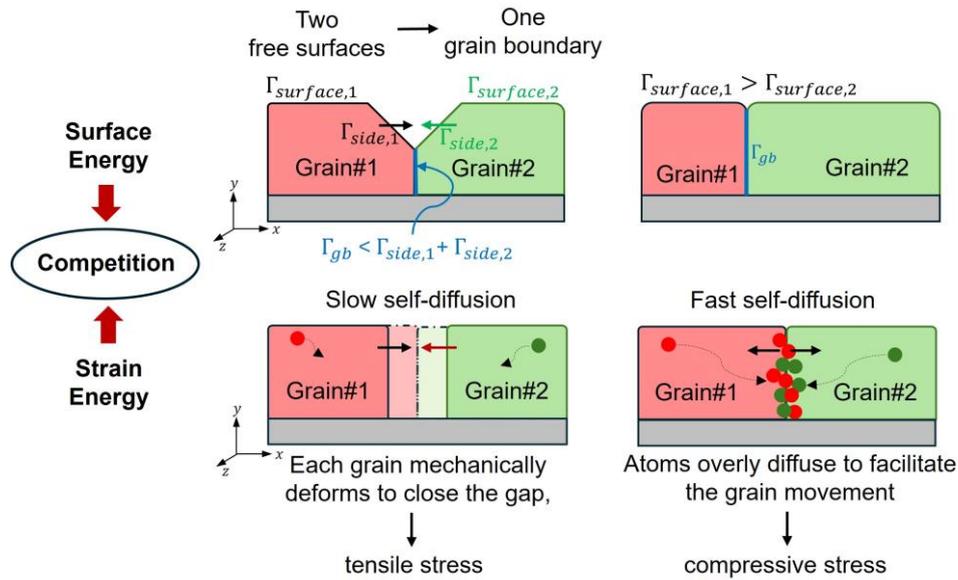

**Figure 1. Schematic illustration of the selective grain growth mechanism in electrodeposited soft metal films.** The schematic highlights the thermodynamic driving force to form a grain boundary, and the competition between surface energy and strain energy in governing grain domination.

## Results & Discussion

**Load stress induced selective grain growth**

Differences in grain selection growth arise from the anisotropy in surface energy and diffusion barrier of soft metal atoms, as informed by density functional theory (DFT) calculations.[16,17] These DFT-informed surface energies and self-diffusion barriers serve as inputs to our phase-field model by establishing correlations with the phase-field constants. Specifically, atomic mobility ($L_q$), which governs growth rate of grains, is modeled to increase exponentially as self-diffusion barriers decrease, following the Arrhenius equation. Additionally, the gradient energy coefficient ($\kappa_q$) at the grain surface is set to increase with the increased surface energy. The larger gradient energy results in higher resistance for a grain to grow. (Mathematical relationships are provided in the Experimental Procedures section)

As illustrated in **Fig. S2**, lattice strain can significantly alter the surface energy. The (001) surface exhibits the highest stability under a compression of 3% for Li metal. Notably, the (101) surface displays the greatest variation in surface energy, with changes approximately 0.08 J/m$^2$.[16] This is because the more densely packed (101) surface of the body-centered cubic (BCC) structure shows a clear preference for lattice expansion, whereas the opposite is true for the less densely packed (001) surface. Given the substantial load stress (>10 MPa) applied with SSE and the soft nature of Li metal (with compressive strength approximately <1 MPa), load stress effects on surface energy have the potential to significantly alter growth behavior. Besides incorporating grain-dependent surface energy and diffusion barrier, phase field simulation was set up with the consideration of deposition rate, areal capacity, deposition temperature, applied stress (more details are provided in the Experimental Procedures section). The simulation started with nucleation at the bottom of the computational domain (simulation snapshot at $t = 0$ in **Fig. 2**). Considering the polycrystalline nature of the Cu substrate, all the nuclei were randomly assigned one of the three representative orientations, namely the (001), (101) or (111). A total of 75 grains were nucleated for statistical significance.

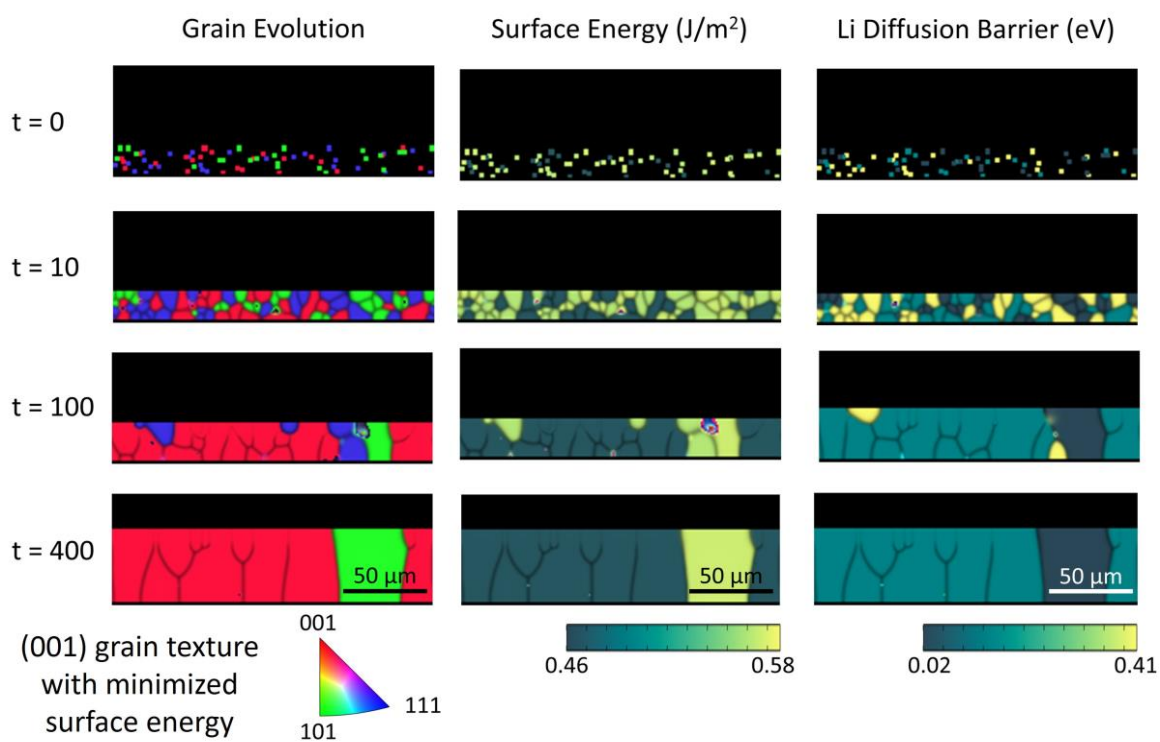

**Figure 2. Grain selection growth for Li metal through thermodynamic theory-based phase-field modeling.** Phase-field simulations showing the grain evolution during electroplating of Li on a Cu substrate with the SSE, together with surface energy and Li diffusion barrier of each grain.

The discrepancy in surface energy is evident at time t = 10 s (refer to the surface energy column in **Fig. 2**). As grains continue to grow and Li deposition progresses, grains with high surface energy, such as (101), are consumed by more thermodynamics favorable grains, notably (001), at time t = 400 s. The substantial variation of surface energy among all the grains outweighs the effects of Li diffusion, leading to grain selection based on minimizing the surface energy. Large load stress applied in the solid state battery helps ensure intimate contact between the SSE and the electrodes, promoting efficient Li ion transport and minimizing interfacial resistance.[18,19] However, induced lattice strain may favor the growth of (001) grain, characterized by a larger lithium diffusion barrier (0.14 eV). This surface energy anisotropy indicates the kinetic constraints of lithium metal anodes with SSEs.

In battery systems utilizing liquid electrolytes, the applied load stress is considerably lower (on the order of hundreds of kPa), resulting in negligible lattice strain on Li metal.[2] As shown in **Fig. S3**, in the liquid electrolyte case, (101) grains are predicted to prevail. This dominance arises due to the similarity in surface energy among all grain orientations, coupled with pronounced discrepancies in lithium diffusion barriers among these orientations. Consequently, selection is based on the Li diffusion barrier, favoring (101) grains which exhibit the lowest Li diffusion barrier (0.02 eV). The anticipated (101) grain texture during lithium electrodeposition with liquid electrolytes has been observed by recent research through X-ray diffraction (XRD) and pole-figure analysis.[20,21] This preference for (101) Li metal texture also elucidates the higher critical current density reported in lithium metal batteries employing liquid electrolytes at room temperature.

Notably, a direct comparison across different alkali metals reveals the surface energy anisotropy of Na and K showing less susceptibility to lattice strain.[16] This suggests that diffusion-favoring (101) grains,[17] characterized by the lowest diffusion barrier of Na (0.04 eV) and K (0.02 eV), may dominate in Na and K metal solid-state batteries operating at room temperature, even under significant load stress (**Fig. S4**). The phase-field simulation for Na and K growth is conducted in a similar manner to the Li growth. As expected, the results show (101) orientation is predominated, and the grain selection is based on maximizing the Na and K diffusion (**Fig. S5** and **S6**). This prediction is supported by a recent experimental result that Na grains in the (101) orientation show preferential growth during deposition with SSEs at 25 °C.[22] These results suggest that Na metal is promising for solid-state batteries. Anode-free solid-state batteries featuring Na metal and bare Al current collectors have recently achieved a critical current density of 1 mA/cm² with reversible cycling, reaching capacities as high as 7 mAh/cm² under a stacking pressure of 10 MPa.[23]

**Temperature effects on grain selection growth**

This section aims to analyze the temperature effects on grain selections. Since (111) grain has the largest diffusion barrier as well as surface energy, indicating the least favorable grain, the competition between (001) and (101) is analyzed. The grain selection can be based on either minimizing the strain energy due to Li diffusion ($\Delta F_{strain}$) or surface energy ($\Delta \Gamma_{surface}$). To determine which mechanism is dominated, the magnitude of each energy term must be calculated based on Eq. S3 and S10 in the Experimental Procedures section. Generally, $\Delta F_{strain}$ is positive and rises as temperature increases. This aligns with the physical perspective that Li atoms exhibit increased mobility at higher temperatures, and thus reduced the strain energy density, potentially making that high mobility grain becomes even more energetically preferable.

By analyzing Li (001) and (110) grains, we use DFT-obtained data and parameters listed in **Table S2**. For the liquid case (**Fig. 3A**), the $\Delta F_{strain}$ is largely positive, while $\Delta \Gamma_{surface}$ is slightly negative, indicating that grain selection is predominantly driven by Li atom diffusion favoring the (101) texture. In contrast, for the solid case (**Fig. 3A**), $\Delta F_{strain}$ and $\Delta \Gamma_{surface}$ are comparable. At room temperature or below, the total energy $\Delta U$ is negative, indicating that (001) grains are favored as controlled by the anisotropy in surface energy. However, as temperature increases, $\Delta F_{strain}$ becomes more significant, outweighing the magnitude of $\Delta \Gamma_{surface}$. As a result, the selection growth of (101) grains becomes possible at high temperatures for the solid case.

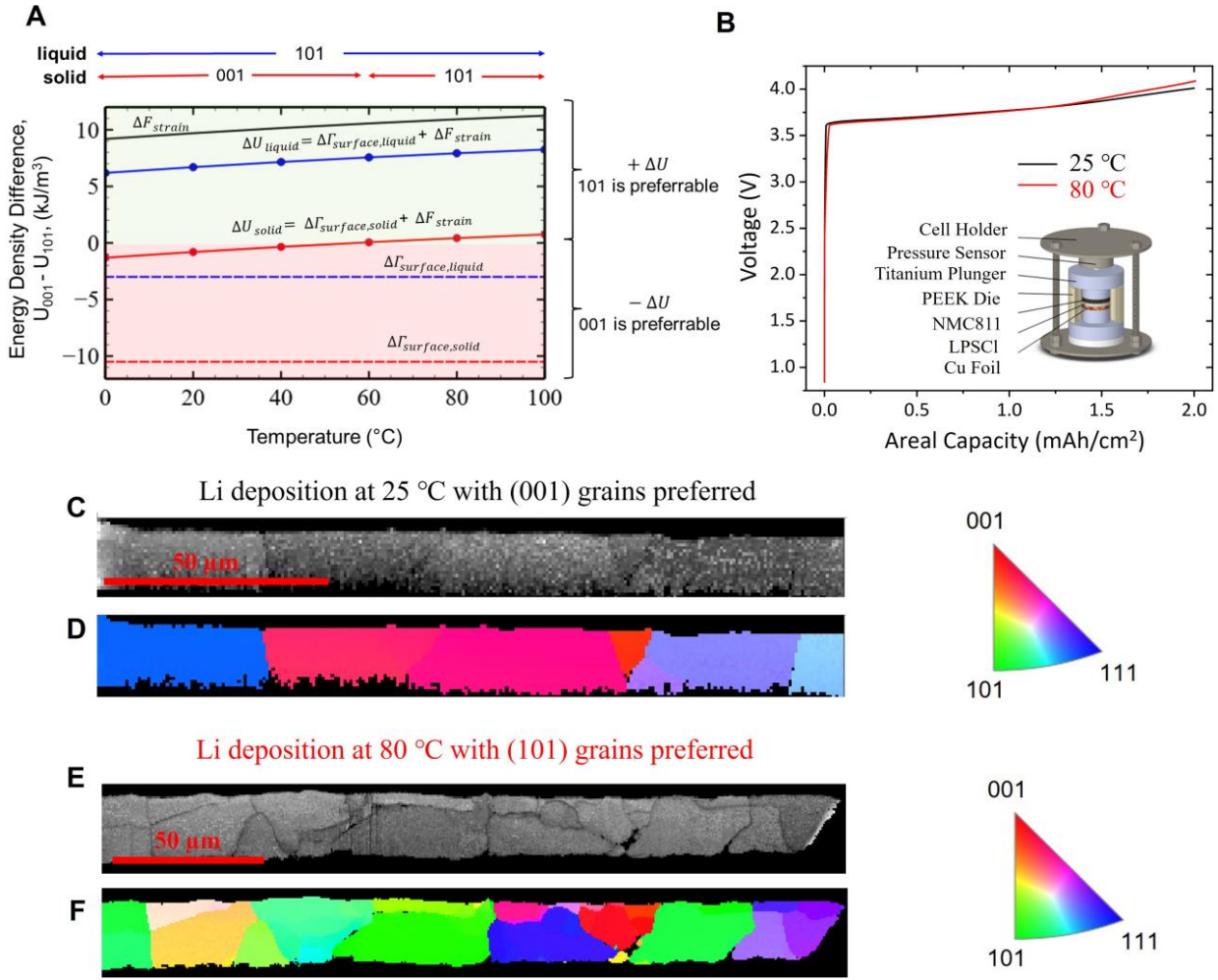

**Figure 3. Temperature effects on grain selection growth of Li metal in anode-free solid-state battery.** (**A**) Analyzing the competition between strain and surface energy density at various temperatures for (001) and (101) grains of Li. When $\Delta U = U_{001} - U_{101} = \Delta \Gamma_{surface} + \Delta F_{strain} > 0$, (101) grain is preferable; if negative, (001) is favored. Temperature influences strain energy change via diffusion rate, with no impact on the surface energy density. (**B**) Voltage profiles during the first Li metal deposition of anode-free solid-state batteries at different temperatures. The battery configuration schematic is shown as the inset. Deposition current density is 0.1 mA/cm$^2$. (**C, D**) The band contrast image and EBSD mapping results along the growth direction for the Li metal deposited at 25 °C. (**E, F**) The band contrast image and EBSD mapping results along the growth direction for the Li metal deposited at 80 °C.

To validate temperature effects in the solid case, Li metal was electrodeposited onto a Cu substrate using a full cell setup, as illustrated in the inset of **Fig. 3B**. A current density of 0.1 mA/cm$^2$ was applied for the Li deposition (2 mAh/cm$^2$) at both 25 °C and 80 °C, with a controlled stacking pressure of 5

MPa. The selection of a relatively small current density was to minimize the current focusing effects (localized Li-ion concentration gradient), which can arise due to surface roughness or defects. In addition, the low current density implies a small Li atom flux to the electrode, keeping the system close to equilibrium, which aligns with the capability of our grain growth phase-field model. At this current density, comparable electrochemical performance is expected for both temperatures, as demonstrated in **Fig. 3B**. To investigate the morphology and texture of the deposited Li, plasma focused ion beam (PFIB) milling coupled with electron backscattered diffraction (EBSD) mapping was employed. PFIB does not require cryogenic temperatures because of the minimal reaction between the $Xe^+$ plasma beam and Li. **Figure S7** displays scanning electron microscopy (SEM) images of deposited Li after PFIB milling. The Li electrodeposition exhibits fully dense morphology for both temperatures investigated. Near the Li/Cu interface, slight porous regions are discernible in the sample deposited at 25 °C. This observation results from the well-documented lithophobic nature of the Cu substrate.[24] As illustrated in **Figure S7B**, the elevated temperature mitigates this issue by improving interface adhesion.

Typically, XRD-based pole-figure analysis was employed to examine the grain orientation growth of Li.[20,21] However, working with Li presents a significant challenge due to its low atomic number. Furthermore, the positioning of Li deposits between a Cu substrate and a thick SSE layer renders XRD impractical for studying the buried thin Li layer (10-20 µm). Little crystallographic characterization of Li electrodeposition in solid-state batteries has been conducted thus far. In this work, PFIB-EBSD with lower acceleration voltage (7 kV) is found to be a suitable approach to provide absolute crystal orientation of every Li grain observed in the cross-section and to create a pole figure for the growth direction (EBSD approach details are provided in the Experimental Procedures section, **Fig. S8** and **Fig. S9**). As shown in **Fig. 3C** and **3D**, the Li grains deposited at 25 °C exhibit preferred growth close to the (001) direction. In contrast, for lithium deposition at 80 °C with the same LPSCl electrolyte, the selection growth of (101) grains becomes apparent (**Fig. 3E** and **3F**). This texture transition aligns with the prediction regarding temperature effects, which is further supported by the statistical analysis on EBSD dataset obtained from three different cross-sections for both temperatures (**Fig. S10** and **Fig. S11**). At room temperature, analysis of Li grains reveals 57.1% with a growth orientation close to (001), followed by 19.0% (101) grains, and 14.3% (111) grains. In contrast, at elevated temperature, the proportion of grains close to (101) increases to 48.7%, those (001) grains decrease to 15.4%. Given that the (101) surface exhibits a smaller in-plane diffusion barrier, surface roughness during the Li stripping process is expected to be better healed by adjacent Li atoms, thereby preventing the formation of whisker-like inactive Li.[25] The uniform deposition and stripping process associated with the Li (101) grains can effectively account for the achievement of current densities exceeding 1.5 mA/cm$^2$ only at elevated temperatures for solid-state batteries featuring a Li metal anode (**Fig. S1**).

**Pressure effects on grain selection growth**

Further parametric studies were performed to generate a phase map for the Li texture growth, considering the anisotropy of surface energy and Li diffusion barrier of the (101) grains which is the lowest among all orientations. As shown in **Fig. 4**, there are two distinct regions: "strain energy minimizing texture", favoring (101), and "surface energy minimizing texture", favoring (001). DFT calculations suggest that increasing pressure raises lattice strain, which in turn alters the surface energy, making it more anisotropic.[16] Once surface energy anisotropy surpasses a critical threshold, grain selection may shift from favoring low strain energy (rapid Li diffusion) to favoring low surface energy grain, particularly under high stacking pressure in solid systems. However, in the liquid systems, where the pressure magnitude is less significant, the system would favor grains with rapid Li diffusion. Furthermore, if the Li diffusion barrier of (101) grains increase and becomes comparable to those of other grains, the anisotropy in Li diffusion is diminished. As a result, (101) grains may become less dominant, potentially leading to an equal mixing with other grains.

Another important factor in the grain growth process, currently not addressed in the model, but worthy of discussion is the pressure effect on the size of the crystal. The pressure applied perpendicular to the surface of the electrodeposited metal could promote in-plane grain coalescence. Given the low yield strength of Li (0.41~0.89 MPa), this effect is particularly pronounced compared to other metals,[26] even at room temperatures. Therefore, it is unsurprising that solid systems subjected to higher stacking pressures exhibit larger sizes in the deposited Li layer. To investigate the pressure effect, we employed a customized load cell (**Fig. S12**) for Li deposition using an ether-based bisalt electrolyte.[27] The selection of a liquid electrolyte aimed to reduce the stacking pressure by at least one order of magnitude for the examination of texture transition. The cross-section morphology at 350 kPa (**Fig. S12**) reveals the formation of columnar Li deposits. Notably, the granular diameter of the deposited Li measures 5-10 μm, considerably smaller than the 30-50 μm observed in the solid-state scenario (**Fig. 3C** and **3E**) at the same current density of 0.1 mA/cm$^2$. Furthermore, the dominant presence of the (101) Li texture is evident in the liquid case, even at room temperature. Prior studies on Li metal texture using various liquid electrolytes have similarly reported the prevalence of the (101) texture,[20,21] suggesting that electrolyte type has minimal influence on selection growth at modest current densities. The existence of the (101) Li texture appears to be an intrinsic characteristic of Li deposition in liquid electrolytes, representing a direct demonstration of the proposed strain energy minimizing texture in the lower pressure region.

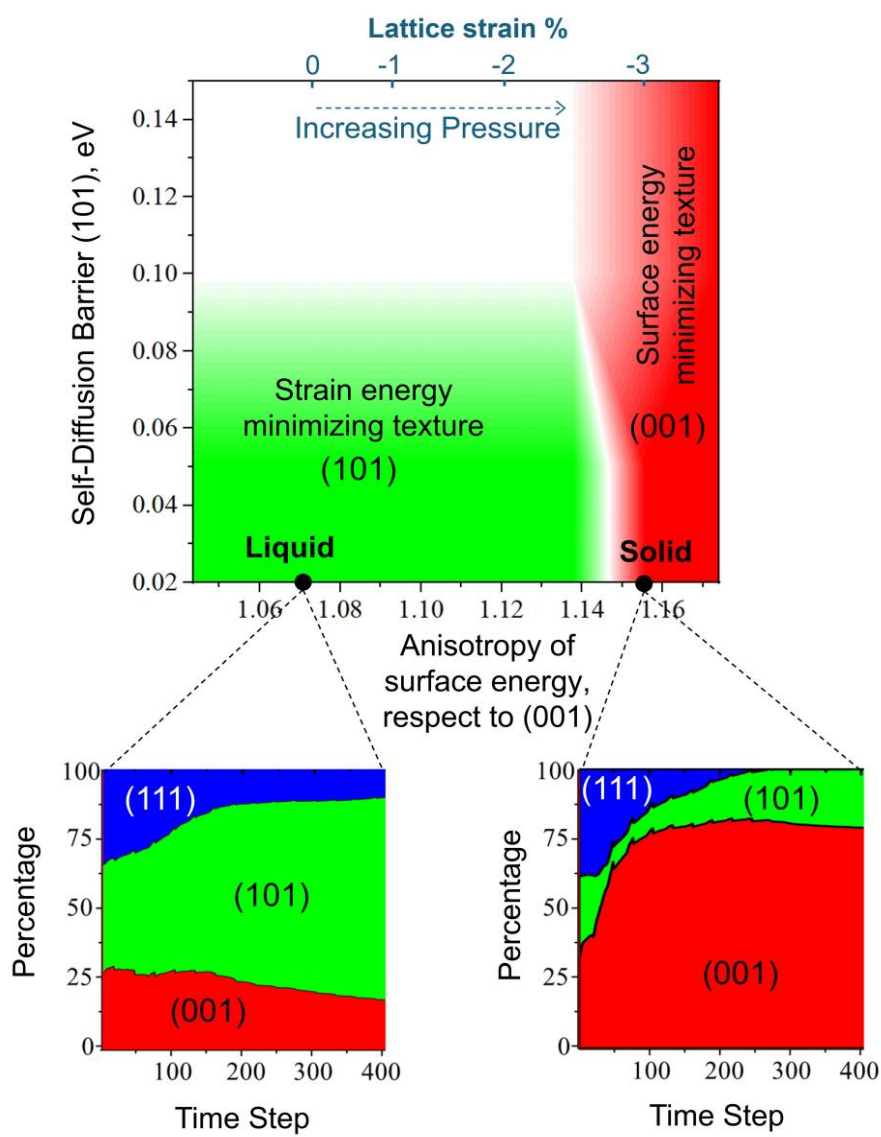

**Figure 4. A phase-map illustrating the phase-field prediction of Li grain selection growth.** Li texture is determined by Li diffusion barrier of (101) grains and surface energy anisotropy, highlighting proportion of each orientation within the simulated liquid and solid system.

**Interfacial layer design for grain selection growth**

Achieving optimal interfacial contact in solid-state batteries requires the application of adequate load stress so that the mechanical properties of the involved solids must be appropriately designed. At any given SSE, a reduced bulk modulus difference between the substrate and Li corresponds to a smaller lattice strain within the Li phase,[28] thereby facilitating the preferential growth of diffusion favoring (101) grains. This effect is pronounced in anode-free solid-state batteries, wherein the current collector functions as the substrate for Li deposition. As shown in **Fig. 5A**, the larger bulk modulus of the Cu substrate ($B$ = 151 GPa) can induce greater lattice strain in the soft Li metal ($B$ = 14 GPa). Previous studies in anode-free solid-state batteries have explored interfacial layer materials (referred to as the "seed" layer), such as Ag and Au, to improve overall performance.[5,29,30] The dotted line connecting pure substance and $Li_xM_{1-x}$ alloy phases in **Fig. 5A** represents a linear relation between the bulk modulus and the Li concentration for Ag and Au seed layer. Despite the inherent lithophilic properties of Ag and Au facilitating the formation of alloy phases with decreased bulk modulus values, achieving a substantial reduction below 30 GPa mandates a considerable lithium alloy concentration (x = 0.8). Motivated by recent DFT calculations,[31] an amorphous Si seed layer to is proposed in this study to reduce the lattice strain within Li metal. In contrast to crystalline structures, amorphous Si and $Li_xSi_{1-x}$ alloy phases exhibit deviations from the anticipated linear relationship between bulk modulus and Li concentration. Consequently, a substantial softening occurs, yielding a bulk modulus below 30 GPa when x exceeds 0.5. Furthermore, with increasing Li content, the bandgap of the Li-Si alloy diminishes, transitioning towards a metallic character suitable for use as current collectors.

A 500 nm-thick layer of amorphous Si (**Fig. S13**) was deposited onto a Cu substrate using the same sputtering technique outlined in our prior investigation.[32] Subsequently, anode-free solid-state full cells were assembled to assess the differences in rate performance between the bare Cu current collector and the Cu current collector with Si deposition. Note the areal capacity for this rate performance assessment was increased to 3 mAh/cm$^2$ to align with application considerations. As shown in **Fig. 5B**, it is evident that the cell with bare Cu experienced a short circuit during the second cycle, even at the low rate of C/40. As for the Si-deposited Cu (**Fig. 5C**), a distinct two-step voltage profile for the first charge emerged due to Si lithiation preceding Li deposition. A slopy curve was observed up to 3.7 V corresponding to an areal capacity of 0.25 mAh/cm$^2$ for Si lithiation, followed by Li deposition onto the Li$_x$Si$_{1-x}$ alloy (**Fig. S14**). The capacity of Si lithiation was calculated to be 2320 mAh/g, corresponding to Li$_{2.43}$Si or Li$_{0.7}$Si$_{0.3}$, a composition falling short of full lithiation up to Li$_{3.75}$Si or Li$_{0.79}$Si$_{0.21}$, thereby maintaining Si within the amorphous phase range to avoid recrystallization.[33] Furthermore, the absence of Si (de)lithiation features below 3.7 V from the first discharge suggests the enduring presence of the formed Li$_{0.7}$Si$_{0.3}$ seed layer throughout subsequent cycling. The anode-free battery featuring Si-deposited Cu sustains operations up to a higher rate of C/2 (1.5 mA/cm$^2$), outperforming other literatures' results at room temperature (**Table S3**). By pairing with the thick cathode, an areal capacity of 9 mAh/cm$^2$ is achieved for the anode-free battery due to the designed Si seed layer (**Fig. S15** and **Fig. S16**).

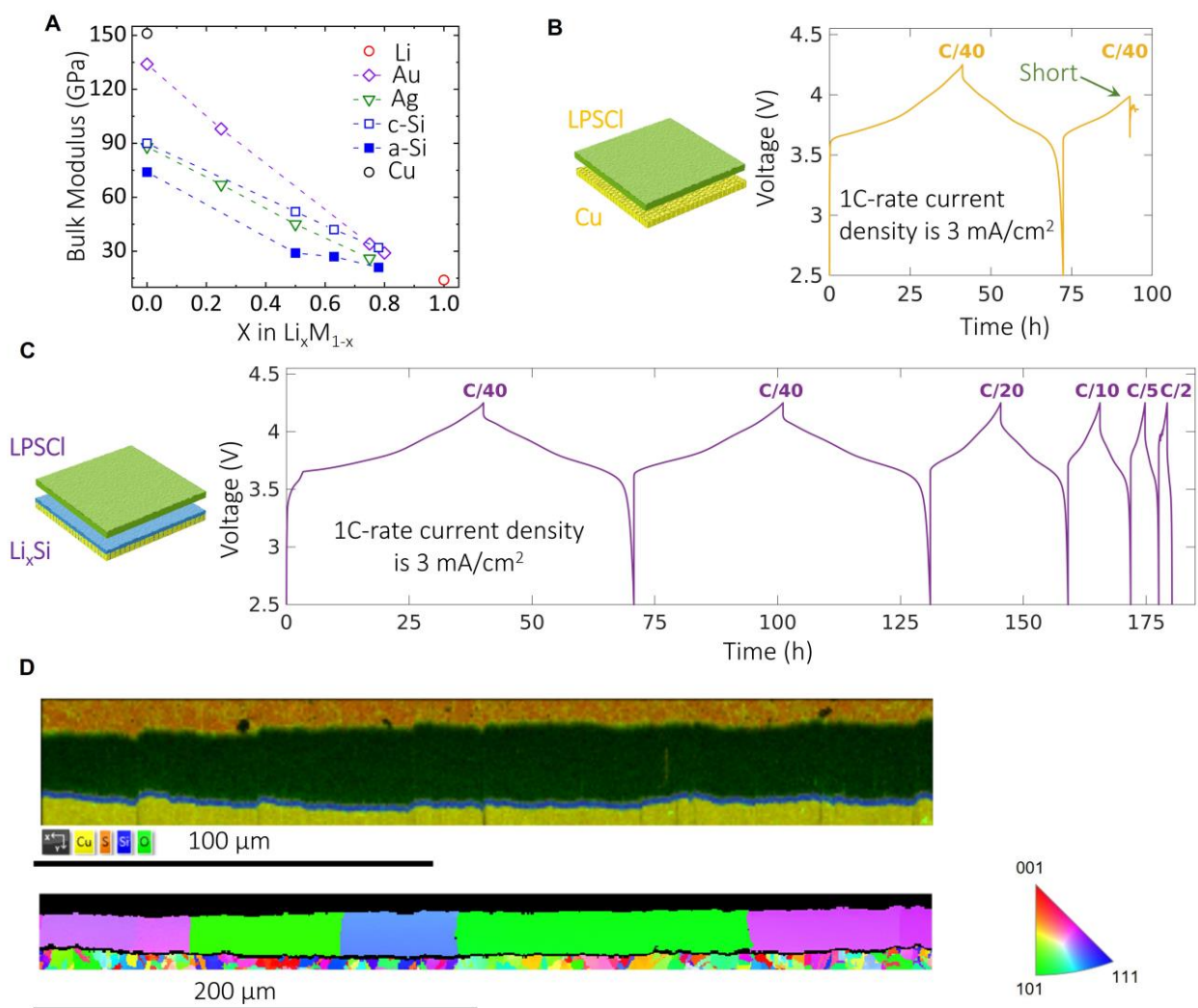

**Figure 5. Interfacial layer design for grain selection growth of Li metal anode in solid-state batteries. (A)** Bulk modulus of Li metal, Cu metal, and different lithium alloys. The data for crystalline and amorphous Li$_x$Si alloys were obtained from reference.[31] The bulk modulus data for Li, Cu, Li$_x$Au, and Li$_x$Ag were sourced from the Materials Project. **(B)** Electrochemical performance of anode-free solid-state full cell with bare Cu as the current collector at 25 °C. **(C)** Electrochemical performance of anode-free solid-state full cell with amorphous Li$_x$Si as the seed layer at 25 °C. **(D)** The EDS and EBSD mapping results along the growth direction for the Li metal deposited with the amorphous Li$_x$Si seed layer at 25 °C. Deposition current density is 0.1 mA/cm$^2$.

To validate the hypothesis regarding grain selection growth, the cross-section of Li metal deposited on the Li$_{0.7}$Si$_{0.3}$ seed layer for an areal capacity of 2 mAh/cm$^2$ was obtained with PFIB for further investigation. Elemental analysis conducted via energy dispersive spectroscopy (EDS) in **Fig. 5D** illustrates the uniform growth of Li between the seed layer and LPSCl electrolyte. The Si seed layer retains its dense and thin film nature, with the thickness increasing to approximately 1.2 μm due to the lithiation process (**Fig. S17**), consistent with findings from prior research.[34] Significantly, even at a deposition temperature of 25 °C, the Li (101) grains become evident with a proportion of 50% with the Si seed layer (**Fig. 5D** and **Fig. S18**). This contrasts with the (001) preferred grains observed previously at the same temperature when utilizing a bare Cu substrate, which manifests the critical role of grain selection growth in facilitating fast kinetics during lithium deposition and stripping processes.

## Conclusions

The intricate electro-chemo-mechanical dynamics inherent in solid-state batteries necessitate model-informed experiments to establish a framework for predictive analysis. This work unveils how the surface energy anisotropy of soft metals can dominate grain selection growth during electrochemical processes, imposing kinetic constraints particularly at room temperature. Leveraging this mechanistic understanding, the critical current density of anode-free solid-state batteries can be improved through the design of a $Li_xSi_{1-x}$ (0.50<x<0.79) interfacial layer. The optimal interfacial layer should meet the following criteria: 1) possess a bulk modulus (<30 GPa) similar to that of Li metal to alleviate load stress induced surface energy anisotropy; 2) exhibit an amorphous structure free of grain boundaries to suppress Li penetration; 3) demonstrate electrochemical stability upon contact with Li metal under reductive potential; 4) offer electronic conductivity to function as the current collector; and 5) display lithophilic properties to lower the nucleation barrier. These findings extend to other soft metal systems, such as Na which exhibits a bulk modulus (8 GPa) even smaller than that of Li. The less surface energy anisotropy of Na can inform the grain selection is based on maximizing the Na diffusion even at room temperature, indicating that Na metal holds promise for solid-state batteries.

# Experimental Procedures

## Resource availability

*Lead contact*

Further information and requests for resources should be directed to and will be fulfilled by the lead contact, Ying Shirley Meng (shirleymeng@uchicago.edu).

*Materials availability*

This study did not generate new, unique reagents.

*Data and code availability*

Request for the data and analysis codes utilized in this work will be handled by the lead contact, Ying Shirley Meng (shirleymeng@uchicago.edu). Requests about the phase-field modeling code utilized in this work will be handled by Lei Chen (leichn@umich.edu).

## Thermodynamics of grain selection growth

The total energy of each grain in a system contains surface, strain and thermal energy density,

$$U = \Gamma_{surface} + F_{strain} + F_{thermal} \tag{S1}$$

When considering two adjacent grains denoted as grain 1 and 2, the analysis for grain selection can be done based on the grain energy density differences:

$$\Delta U_{12} = U_1 - U_2 = \Delta\Gamma_{surface} + \Delta F_{strain} + \Delta F_{thermal} \tag{S2}$$

When $\Delta U_{12}$ is negative, grain 1 is more energetically preferable than grain 2, and vice versa. Furthermore, due to the thin nature of deposited film (< 50 µm), heat generated during deposition is expected to efficiently dissipate to the surface. Therefore, stored thermal energy is assumed isotropic for all orientations, and thermal contribution to the grain selection can be negligible, $\Delta F_{thermal} \sim 0$. In addition, surface energy difference can be written as:

$$\Delta\Gamma_{surface} = (E_{S,1} - E_{S,2})/h \tag{S3}$$

where $h$ is the film thickness, and $E_{S,i}$ is the surface energy of grain $i$. Noticeably, the magnitude of $\Delta\Gamma_{surface}$ inversely depends on the film thickness. As the deposited film grows, the effect of surface energy on the grain selection is diminishing. Moreover, the strain energy density is stored mechanical energy due to grain deformation. According to the experimental measurements,[13,14] soft metals exhibit a power-law creep behavior. We thus formulate the strain energy by assuming the linear-elastic perfectly-plastic curve which can be written as:

$$F_{strain} = F_{elastic} + F_{plastic} = \frac{1}{2}E\varepsilon_y^2 + \sigma_y[\varepsilon - \varepsilon_y] \tag{S4}$$

or

$$F_{strain} = \sigma_y\varepsilon - \frac{1}{2}E\varepsilon_y^2 \tag{S5}$$

where $\sigma_y$, $\varepsilon_y$, and $E$ represent yield strength, yield strain, and elastic modulus, respectively. The yield strength may change significantly with size. The smaller size scale leads to higher strength, especially in soft metals like Li. As the film grows, the yield strength tends to decrease.[35] Assuming that yield strength is isotropic, the strain energy density difference between grain 1 and 2 is

$$\Delta F_{strain} = \sigma_y[\varepsilon_1 - \varepsilon_2] - \frac{1}{2}(E_1 - E_2)\varepsilon_y^2 \tag{S6}$$

Since elastic range for soft metals is usually small, the elastic strain energy is negligible. The equation S5 is simply reduced to:

$$\Delta F_{strain} = \sigma_y[\varepsilon_1 - \varepsilon_2] \tag{S7}$$

In addition, total strain in each grain, $\varepsilon_i$, can be decomposed to intrinsic and extrinsic strain.

$$\varepsilon_i = \varepsilon_{intrinsic} + \varepsilon_{extrinsic} \tag{S8}$$

The extrinsic strain can result from external factors such as stacking pressure. However, soft metals typically behave with a creep flow like a fluid under a stack pressure >1 MPa. Under stacking pressure higher than yield strength of soft metals, creep strain follows the power law: $\dot{\varepsilon}_{extrinsic} = A\sigma^n$.[36,37] The creep constants ($A$, $n$) may exhibit some anisotropy, leading to varying strain responses. However, at high applied stress, the effect of creep anisotropy could be small compared to the anisotropy of other factors, such as, surface energy and self-diffusion barriers. For liquid systems with stacking pressures in the hundreds of kPa or lower, creep behavior may shift to diffusional creep,[36] but the resulting extrinsic strain rate is likely minimal and thus excluded from analysis. Meanwhile, the intrinsic strain is deposition-induced strain. Two neighboring grains close the gap between each other to reduce their free-surfaces, by forming a grain boundary. Generally, such a process can be contributed by mechanical work, i.e., the deformation of the grain. Tensile stress or strain is generated within as two grains pulling toward each other. However, the diffusivity of atoms may alleviate the built-up tensile stress, which can be thought of as a mechanism similar to diffusional creep. Through atomic movement, this mechanism provides strain energy relief, minimizing strain energy accumulation during grain boundary formation. The relationship between strain and atomic mobility is

$$\varepsilon_{intrinsic} = \varepsilon_c + (\varepsilon_T - \varepsilon_c)exp\left(\frac{-\beta D_i}{LR}\right) \tag{S9}$$

where $D_i$ is the atomic diffusion coefficient of each grain. $\varepsilon_T$ is the tensile strain induced when grains mechanically deforming to close the gap without atomic diffusion assistance. $\varepsilon_c$ is the compressive strain from atomic additions to the grain boundary, and $\beta$ is a fitting constant. $L$ and $R$ is grain size and the deposition rate, respectively. If atomic diffusion or $D_i$ is very small that $exp\left(\frac{-\beta D_i}{LR}\right) \sim 1$, tensile strain is expected. If atomic diffusion is large, $exp\left(\frac{-\beta D_i}{LR}\right) \sim 0$, compression is predicted. Combining Eq. S5, S6, and S7, one can write:

$$\Delta F_{strain} = \sigma_y[\varepsilon_1 - \varepsilon_2] = A\left[exp\left(\frac{-\beta D_1}{LR}\right) - exp\left(\frac{-\beta D_2}{LR}\right)\right] \tag{S10}$$

where $A$ is a constant expressed as, $A = \sigma_y(\varepsilon_T - \varepsilon_c)$. If $D_1 > D_2$, $\Delta F_{strain}$ is negative, giving grain 1 is preferable than grain 2. Furthermore, the diffusion coefficients can be correlated with diffusion barrier $E_{a,i}$, through Arrhenius equation, written as

$$D_i = D_0 exp\left(-\frac{E_{a,i}}{k_B T}\right) \tag{S11}$$

where $D_0$ is a diffusion constant, $k_B$ is Boltzmann constant, and $T$ is temperature.

**Temperature analysis for grain selection growth**

General idea is to compare the magnitude in each type of energy density differences between (001) and (101) grains, $\Delta F_{strain}$ and $\Delta \Gamma_{surface}$. If $\Delta F_{strain} + \Delta \Gamma_{surface}$ is positive, (101) grains are more energetically favorable; otherwise, (001) grains are preferred. Grain size ($L$) and thickness ($h$) are set to be 10 μm, as this scale matches the grain morphology observed experimentally. The deposition rate ($R$) is 0.001 μm/s, approximated from the C-rate and areal capacity applied in the experiment. Yield strength ($\sigma_y$) is assumed to be 0.55 MPa within the measured range.[13] The maximum possible tensile and compressive strain are assumed to be 0.2, given that lithium can undergo significant deformation due to creep. Pre-factor for diffusion ($D_0$) is 1 x $10^{-15}$ m$^2$/s, which is in the typical range of reported Li self-diffusion.[38] Lastly, $\beta$ is a fitting constant, which varies between materials and requires experimental calibration. Measuring stress and strain to calibrate $\beta$ in the deposited film is beyond the scope of this work. Therefore, $\beta$ is assumed to be 1, which does not affect the trend of this analysis.

**Phase-field model for grain selection growth**

The typical phase-field model for multiphase grain nucleation and growth has been extended to simulate the grain selection in the soft metal electrodeposition under different environments. Phase variables, $(\phi_1, \phi_2, ..., \phi_n)$, are introduced to associated with each grain in the system, total number of n. Grains evolve to reduce the overall free energy in the system, which can mathematically be written, following the Allen-Cahn equation[39] as:

$$\frac{\partial \phi_q}{\partial t} = -L_q(T) \frac{\partial F}{\partial \phi_q} \tag{S12}$$

where $q = 1, 2, ..., n$ and $L_q(T)$ is the temperature-dependent phase-field parameter related to the mobility of atoms. Grains with faster atomic mobility allow for quicker hopping of atoms across the grain boundaries, reducing the overall strain energy. The contribution of strain energy can thus be reflected in the growth mobility $L_q(T)$. $F$ is the free energy functional, expressed as

$$F = \int \left[ w \cdot f_0(\phi_1, \phi_2, ..., \phi_n) + \sum_{q=1}^{n} \frac{\kappa_q}{2} (\nabla \phi_q)^2 \right] d\vec{r} \tag{S13}$$

The first term on the right-hand side is the local free energy density, $f_0(\phi_1, \phi_2, ..., \phi_n)$ follows Landau expressions. The second term represents grain surface energy, which can be thought of as the resistance to development of the grain boundary. The resistance to growth for each grain depends on a constant, $\kappa_q$. In most cases, phase-field constants, $L_q$ and $\kappa_q$, take normalized values computed from physical properties; therefore, the correlation to the physical system as well as first-principle calculations are required.

**Bridging DFT information to phase-field modeling constants**

From a physical aspect, the growth rate of each grain can be understood in terms of atom diffusion. Grains with faster atomic diffusion allow for quicker hopping of atoms across the grain boundaries, reducing the overall strain energy – a phenomenon aligns with kinetics considerations. Nevertheless, grains with higher surface energy are energetically less favorable, suppressing the growth rate and even being consumed by surrounding grains as the system tends to minimize the overall energy, reflecting thermodynamics considerations.

We propose that atom diffusion is associated with $L_q$, a temperature-dependent mobility term in the phase-field model, which is correlated to the atom diffusion barrier, $E_a$ (eV), through Arrhenius

equation, written as:

$$L_q = L_1 exp\left(L_2 \frac{-E_a}{k_B T}\right) \quad (S14)$$

where $L_1$ is a possible maximum atomic diffusion, which could be different from one to another system depending on the surface chemistry. $L_2$ is a fitting constant, and $T$ is temperature. With a larger value of $L_q$, rapid movement of a particular grain boundary is expected, resulting in a relatively large grain size compared to its neighboring grains.

Meanwhile, the surface energy is directly associated with gradient energy coefficient term, through an exponential equation as follows:

$$\kappa_q = \kappa_1 exp\left(\kappa_2 \frac{E_s}{E_0}\right) \quad (S15)$$

where $\kappa_1$ and $\kappa_2$ are fitting constants, and $E_0$ is the reference grain surface energy, by which assumes the average $E_s$ among the grains. Higher surface energy corresponds to a higher $\kappa_q$, resisting grain growth. Note that this term related to surface energy is not temperature-dependent by assuming surface remains solid far below the melting point.

In literature, DFT data including the atom diffusion barrier, and surface energy for each grain orientation is available, as plotted in **Fig. S4**. The fitting parameters in Eq. S14 and S15 were selected such that the magnitudes for $L_q$ and $\kappa_q$ fall within the range of 0.1-10, which is not significantly far from their nominal values. This range allows for the observation of the interplay between the two parameters without interrupting simulation convergency for all the three soft metals (Li, Na, and K): $L_1 = 7.5$; $L_2 = 0.27$; $\kappa_1 = 4.1 \times 10^{-9}$, $\kappa_2 = 20$. Taking the Li case as an example, the conversion from DFT derived data to phase-field input parameters is illustrated in **Fig. S2**.

**Phase-field simulation setup**

The nucleation site and type of grains are assumed identical for both liquid and solid cases, to illustrate the effect of grain domination. Periodic boundaries were then applied on the left and right sides of the computational domain, while zero flux boundary condition was implemented at the bottom and top boundaries.

Upon grain nucleation and boundary setup, the competitive growth throughout the progressive deposition was enabled by carrying out grain growth in a layer-by-layer fashion. This was done by dynamically activating the computational domain in an incremental, layer-wise manner. Without doing that and due to the inherent limitation of phase field model, the grain would grow upwards aggressively and occupy the entire domain instantly, thus deviating from the gradual deposition with sufficient grain competition in practice.

The above-customized grain growth phase field model was solved using finite difference method with explicit Euler scheme. All the simulations were performed in a rectangular domain with 200 $\Delta x$ × 70 $\Delta x$ grids, where $\Delta x$ is the grid size and was chosen as 0.5 μm. The time step of $\Delta t = 0.01$ s was used for time discretization. Initial grain nucleation is set to random spots with random orientations (represented by colors). During the simulation, grains grow and occupy more black space in the domain.

**PFIB milling and SEM**

The Li metal deposition samples were carefully stored in an Argon environment prior to microscopy analysis. To prevent oxidation and reaction with the atmosphere, all the samples were swiftly loaded into the microscope chamber.

For the analysis, we utilized the Thermo Scientific Helios 4 UXe PFIB, which allowed for large area cross-sectional milling, as well as subsequent EBSD and EDS analysis. The solid and liquid cell samples were mounted on a specialized 45-degree pre-tilted holder. $Xe^+$ ions were employed to mill and clean the cross-sections of various electrode samples.

To generate a cross-section that through the entire thickness of the electrode, including the Li metal and the underlying Cu foil, rough milling was performed at a FIB acceleration voltage of 30 kV and beam current of 2.5 μA, resulting in a cross-section width of up to 900 μm. This rough milling step also served to eliminate any surface oxidation that might have occurred during the brief exposure of the samples to the atmosphere during the loading process. Following rough milling, the electrode cross-sections were cleaned using a beam current of 200 nA. Both the intermediate and final cleaning were carried out at 30 kV.

**EBSD and EDS analysis**

EBSD patterns and maps of the Li metal were collected using an Oxford Instrument Symmetry EBSD camera. The SEM conditions for EBSD collection were set at an acceleration voltage of 7 kV and a beam current of 6.4 nA. The Li metal was first indexed using the default lithium phase in the Oxford Aztec software. Additionally, EDS maps of the same region were obtained using an Oxford Instrument Ultim Max 170mm$^2$ detector. All EDS maps are "True Maps": after peak deconvolution and background correction.

Due to Li low atomic number and weak interaction with the electron beam, a 7 kV beam was used for the EBSD mapping, and 3 patterns were averaged for each pixel. Each data point (or pixel) on the EBSD mapping corresponds to a single scan spot on the sample. The colors in the image represent different crystallographic orientations, where each color in the mapping reflects a specific orientation family. Adjacent pixels with different colors indicate boundaries between regions with different orientations, which can correspond to grain boundaries. **Fig. S8** demonstrates the importance of using lower voltage to minimize the interaction volume with Li. To minimize beam damage during acquisition the pixel sampling was set to 0.75 μm. Even using these conditions, the resulting bands are few and weak. As a result, significant regions in the map produced a low number of weak bands that were not solved using the default setup using the Hough transform. To overcome this, a pattern matching approach was applied, using "Mapsweeper", a software package offered by Oxford Inst. One big advantage of the method is that one can verify the solution for each pattern is correct. **Fig. S9** provides an example of such a solution for an originally unsolved pattern together with the dynamical diffraction simulation and the pattern match quality. The analysis was done using only the refinement and repair options and a minimum cross-correlation factor of 0.15 between the measured and simulated EBSD pattern.

**Full cells assembling and electrochemical measurements**

Li$_6$PS$_5$Cl (LPSCl) from NEI Corporation (USA) and Mitsui Kinzoku (Japan) was used as both catholyte and solid-state electrolyte separator. LiNi$_{0.8}$Mn$_{0.1}$Co$_{0.1}$O$_2$ (NMC811) with a boron-based surface coating from LG Chem (Republic of Korea) was selected as the cathode. As a conducting agent, vapor grown carbon fiber (VGCF) from Sigma Aldrich (Graphitized, Iron-free) was vacuum dried overnight at 160 °C to remove moisture. Cathode composite was hand-mixed with a weight ratio of NMC811 : LPSCl : VGCF = 66 : 31 : 3. The custom-made solid-state battery pellet cell made of two titanium rod current collector and 10 mm inner diameter polyether ether ketone (PEEK) holder was used for anode-free solid-state cell cycling. The 75 mg of LPSCl was compressed at 370 MPa as a solid-state separator. Afterwards, Cu foil or Si deposited Cu foil and NMC cathode composite (2~3 mAh/cm$^2$ loading) were inserted to each end of the separator layer, and then pressed at 370 MPa. The as-fabricated cell was cycled with the stack pressure of 5 MPa and in a range of temperature from 25 °C to 80°C.

## Acknowledgements

This work is funded by the Energy Storage Research Alliance "ESRA" (DE-AC02-06CH11357), an Energy Innovation Hub funded by the U.S. Department of Energy, Office of Science, Basic Energy Sciences. PFIB data collection at Americas NanoPort was supported by the funding and collaboration agreement between UCSD and Thermo Fisher Scientific on Advanced Characterization of Energy Materials. K. Tantratian, Z. Wang and L. Chen are supported by the National Science Foundation under Grants No. 2323475 and UM-Dearborn start-up. Part of the PFIB experiments were conducted at the University of Southern California in the Core Center of Excellence in Nano Imaging. Part of the FIB-SEM in this work was performed at the San Diego Nanotechnology Infrastructure (SDNI) of the University of California San Diego, a member of the National Nanotechnology Coordinated Infrastructure, which is supported by the National Science Foundation (grant ECCS-2025752). M. Z. thanks Diyi Cheng, Bingyu Lu, and Yu-Ting Chen for their assistance with the sample preparations. M. Z. thanks Brandon van Leer and Ken Wu for their assistance with PFIB data collection and analysis.

## Author contributions

M. Z. conceived the project. M. Z., L. C., and Y. S. M. supervised the project. M. Z., K. T., Z. W., M. C., L. C., and Y. S. M. performed the mechanism and modeling analysis. M. Z., Z. L., L. L., and A. A. performed PFIB-EDS-EBSD experiments and analysis. S. Y. H., R. S., and S. B. prepared samples for PFIB-EBSD analysis. S. Y. H. conducted electrochemical testing for anode-free solid-state battery. M. Z. prepared the initial draft of the manuscript. M. Z., K. T., Z. W., M. C., S. Y. H., R. S., L. C., and Y. S. M. organized the work and helped with the draft of the manuscript. All the authors discussed the results and approved the final version of the manuscript.

## Declaration of Interests

## Inclusion and Diversity

We support inclusive, diverse, and equitable conduct of research.